\documentstyle{mn}
\input{epsf}

\newcommand{\bez}{\begin{eqnarray*}}
\newcommand{\eez}{\end{eqnarray*}}
\newcommand{\be}{\begin{equation}}
\newcommand{\ee}{\end{equation}}
\newcommand{\beq}{\begin{eqnarray}}
\newcommand{\eeq}{\end{eqnarray}}
\newcommand{\bc}{\begin{center}}
\newcommand{\ec}{\end{center}}

\newbox\grsign \setbox\grsign=\hbox{$>$} \newdimen\grdimen \grdimen=\ht\grsign
\newbox\simlessbox \newbox\simgreatbox \newbox\simpropbox
\setbox\simgreatbox=\hbox{\raise.5ex\hbox{$>$}\llap
     {\lower.5ex\hbox{$\sim$}}}\ht1=\grdimen\dp1=0pt
\setbox\simlessbox=\hbox{\raise.5ex\hbox{$<$}\llap
     {\lower.5ex\hbox{$\sim$}}}\ht2=\grdimen\dp2=0pt
\setbox\simpropbox=\hbox{\raise.5ex\hbox{$\propto$}\llap
     {\lower.5ex\hbox{$\sim$}}}\ht2=\grdimen\dp2=0pt
\def\simgreat{\mathrel{\copy\simgreatbox}}
\def\simless{\mathrel{\copy\simlessbox}}

\def\Ls{L_{s}^{\rm obs}}
\def\Lh{L_{h}^{\rm obs}}
\def\Lsintr{L_s^{\rm intr}}
\def\Lstot{L_{s}^{\rm tot}}
\def\fcol{f_{\rm col}}
\def\fGR{f_{\rm GR}}
\def\taut{\tau_T}
\def\taup{\tau_p}
\def\Te{T_e}
\def\Ts{T_s}
\def\sigmat{\sigma_T}

\begin{document}

\title[Spectral Transitions in Accreting Black Holes] 
{The Nature of Spectral Transitions in Accreting Black Holes: The Case
of Cyg X-1} 

\author[J. Poutanen, J.~H. Krolik and F. Ryde]
{\parbox[]{6.8in} {Juri Poutanen$^{1,2\star}$, Julian H. Krolik$^{3\star}$,
and Felix Ryde$^{2\star}$} \\
$^{1}$Uppsala Observatory, Box 515, S-751 20 Uppsala, Sweden; 
juri@astro.uu.se \\
$^{2}$Stockholm Observatory, S-133 36 Saltsj\"obaden, Sweden \\
$^{3}$Johns Hopkins University, Baltimore, MD 21218, USA} 

\date{Accepted, Received}

\maketitle


\begin{abstract}
Accreting black holes radiate in one of several spectral states, switching
from one to another for reasons that are as yet not understood.  Using
the best studied example, Cyg X-1, we identify the geometry and physical
conditions characterizing these states.  In particular, we show that in the
hard state most of the accretion energy is dissipated in a corona-like
structure which fills the inner few tens of gravitational radii around the
black hole and has Compton optical depth of order unity.  In this
state, an optically thick accretion disc extends out to greater distance,
but penetrates only a short way into the coronal region.
In the soft state, the optically thick disc moves inward and receives the
majority of the dissipated energy, while the ``corona" becomes optically
thin and extends around much of the inner disc.  The mass accretion rate
in both states is $\sim 1 \times 10^{-8} M_{\odot}$ yr$^{-1}$.
\end{abstract}

\begin{keywords}
{accretion, accretion discs --  gamma-rays: theory -- 
radiation mechanisms: non-thermal -- 
stars: individual (Cygnus X-1) -- 
X-ray: general  -- X-ray: stars}  
\end{keywords}

\section{SPECTRAL STATES OF GBHC} 
\label{sec:intro}

   For many years we have known that Galactic Black Hole Candidates (GBHC)
radiate X-rays in one of several spectral states, switching suddenly from
one to another.  Although intermediate states are sometimes also seen, the
spectral states of GBHCs are best typified by their extremes:
a ``hard" state (sometimes also called
the ``low" state because the 2 -- 10 keV flux is relatively weak in this
state), and a ``soft" state (sometimes also called the ``high" state
because the 2 -- 10 keV flux is relatively strong).  The spectrum in
both states may be roughly described as the sum of two components: a
blackbody, and a power-law with an exponential cut-off.  In the soft state
the blackbody is relatively prominent and the power-law is steep; in
the hard state more energy is carried in the power-law, whose slope is
then shallower.
\footnotetext{$\star$ E-mail: juri@astro.uu.se (JP); 
jhk@gauss.pha.jhu.edu (JHK); felix@astro.su.se (FR)} 

   More specifically, in the soft state (SS), the 0.1 -- 5~keV band is dominated
by a component best described as either a blackbody with temperature
0.3 -- 1~keV, or perhaps a sum of blackbodies all with temperatures in
that range \cite{mi84}.  The slope (in energy units) of the power-law
between 5 and 40 keV is generally 1.0 -- 1.5 \cite{tanaka95}.
OSSE and RXTE data,  when fitted with 
a model consisting of a power-law with an exponential cut-off, 
show slopes between 1.6 and 2.1 and a cut-off  energy 
(although not very well determined)  $\simgreat 200$ keV
(Grove et al. 1997, see  also Phlips et al.1996, Cui et al. 1996 and 
Gierli\'nski et al. 1997b for observations of Cyg X-1).  
Most of the energy is emitted in the soft thermal component.

In the hard state (HS), the soft component is much less luminous, and can be 
represented by a black body with a temperature 0.1-0.2~keV 
\cite{bal95,ebis96}.
Most of the energy is emitted in the hard tail, which can be 
represented by a power-law of slope 0.3 -- 0.7
with an exponential cut-off at about 100 keV 
\cite{tanaka95,phl96,grove97,gier97,zdz97}. 
The {\it Ginga} spectrum of Cyg X-1 also shows 
a clear sign of hardening at about 10 keV accompanied by a
fluorescent Fe K$\alpha$ line at $\sim$ 6.4 keV.  These effects are
most easily interpreted as the signature
of Compton reflection from cold matter in the vicinity of the
hard X/$\gamma$-ray source.  In this interpretation, the strength of
the Fe K$\alpha$ line and the reflection bump
requires the cold matter to subtend a solid angle
$\Omega/2\pi\approx 0.4$ around the hard X-ray source
\cite{ebis96,gier97}.

     It was originally thought that these systems were much more luminous
in the soft state than in the hard.  However, in a striking new result,
Zhang et al. (1997b) have shown that the {\it bolometric} luminosity of Cyg X-1
changes only slightly despite dramatic changes in spectral shape.  In
view of this, we eschew the designations ``high" and ``low", and strongly
urge that in the future these states be called ``soft" and ``hard".

   What exactly is happening in these sources when they switch spectral
state has long been a puzzle.  However, in recent years advances have
been made both on the observational side (most importantly,
hard X-ray/soft $\gamma$-ray observations now give us a clearer picture
of that portion of the spectrum) and on the theoretical side (we now
understand much better how thermal Comptonization works when the seed
photons are produced in large part by reprocessing part of the hard
X-ray output).  Making use of this progress, we can now use the observed
character of the spectrum in these states to
determine the geometry and energy dissipation distribution in an
accreting black hole system in {\it all} its different spectral states.  The
geometry and energy dissipation distribution inferred purely
on the basis of {\it radiation} physics can then be
used to guide efforts to obtain {\it dynamical} explanations for the
changes in spectral state.  Because
the observational data are best for Cyg X-1, we illustrate our method
here by applying it to that source.  As will be seen, even in this
example, there are still uncertainties which prevent some of our conclusions
from being as strong as might be possible in principle, but the bulk of this
program is now realizable in the case of Cyg X-1, and it may soon be
possible to extend it to other sources.

\section{DIRECTLY MEASURED PARAMETERS AND INFERENCES FROM ANALYTIC ARGUMENTS}

     We attribute the two components with which the spectra are fit to
two physically distinct, but related, regions: an optically thick, quasi-thermal
region which is responsible for the blackbody component (the accretion
disc?); and an optically thin very hot region which radiates the hard
X-rays (a corona?).  We call the intrinsic dissipation rates in the
``disc" and ``corona" $\Lsintr$ and $L_h$, respectively.  The scale
over which the ``disc" radiates most of its energy is $R_s$.

    Following the consensus in the field 
\cite{sle76,hmg94,stern95,zdz97}, we suppose that the hard X-rays are produced
by thermal Comptonization.  The seed photons for this process are partly
created locally (by thermal bremsstrahlung or synchro-cyclotron radiation:
Narayan \& Yi 1995)
and are partly created in the quasi-thermal region.  The luminosity of
the quasi-thermal region is in turn partly due to local energy
dissipation, and partly due to reradiation of hard X-rays created in the
``corona".

    In this model, the shape of the Comptonized spectrum produced
by the ``corona" may be quite adequately described phenomenologically
by two parameters: the power-law slope $\alpha$, the exponential cut-off
energy $E_c$. 
Two more parameters (the effective temperature $T_s$ and $\Ls$)
define the soft part of the radiation. The ratio
of the observed hard luminosity to the observed soft luminosity, $\Lh/\Ls$,
and the magnitude of the reflection bump $C\equiv\Omega/2\pi$ 
($C=1$ corresponds to the amplitude of reflection expected from a
slab subtending a  $2\pi$ solid angle around an isotropic X-ray source
atop the slab) complete the set of observables.
These phenomenological parameters are determined
by two dimensional quantities, the total dissipation rate and $R_s$, and 
four dimensionless physical parameters: the ratio $\Lsintr/L_h$;
the Compton optical depth of the ``corona" $\taut$; the fraction $D$ of the 
light emitted by the thermal region
which passes through the ``corona"; and the ratio $S$ of intrinsic seed photon
production in the ``corona" to the seed photon luminosity injected from outside.
Another dimensionless parameter, the compactness 
$l_h \equiv L_h\sigma_T/(m_e c^3 R_h)$, may be used to determine
the relative importance of $e^{\pm}$ pairs in the corona ($R_h$ is the size 
of the corona).  In this
context it is also useful to distinguish the net lepton Compton optical depth
$\taup$ from the total Compton optical depth (including pairs), $\taut$.

   We will show how all these parameters, as well as several others of
physical interest, may be inferred from observable quantities.  The observables
on which we base this analysis for Cyg X-1 are shown in Table 1.  Because
Cyg X-1 varies, the numbers seen at any particular time may be somewhat
different from the ``typical" values we cite.

\begin{table}
\label{tab:1}
\caption{Observational characteristics of Cyg X-1 
in its soft and hard states}
\begin{tabular}{llll}
\hline
{Parameter} & {Hard state} & { } & {Soft state} \\
$\Ls$$^{\;\;\rm a}$  & $1 \times 10^{37}$ erg/s$^{[1,2,3]}$
& &  $4 \times 10^{37}$ erg/s$^{[4,5]}$  \\
$\Lh$$^{\;\;\rm b}$     &   $4\times 10^{37}$ erg/s $^{[4,6]}$ 
&          &    $1 \times 10^{37}$ erg/s $^{[4,5]}$ \\ 
$\alpha$$^{\rm c}$  &   0.6$^{[6]}$  
 &         &     1.6$^{[7,9,12]}$   \\ 
$\Ts$$^{\rm d}$ & 0.13 keV$^{[1,2,3]}$
 &     &     0.4 keV$^{[7,8,9]}$ \\  
$E_c$$^{\rm e}$  &  150 keV$^{[6,10,11]}$ 
 &     &     $\simgreat 200$ keV$^{[7,10]}$  \\ 
$C$$^{\rm f}$     &   0.4$^{[3,6]}$
 &      &     0.55$^{[12]}$        \\ 
\hline
\end{tabular}

$^{\rm a}${Observed soft luminosity
 (total luminosity below $\sim 1$ keV in the HS
and below $\sim 3$ keV in the SS)} \\
$^{\rm b}${Observed hard luminosity} \\
$^{\rm c}${Energy spectral index} \\
$^{\rm d}${Temperature of the soft component}\\
$^{\rm e}${Cut-off energy of hard component} \\
$^{\rm f}${Covering factor of the cold matter}\\
References are:{(1) Ba{\l}uci\'nska \& Hasinger 1991; 
(2) Ba{\l}uci\'nska-Church et al. 1995; 
(3) Ebisawa et al. 1996; 
(4) Tanaka \& Lewin 1995; 
(5) Zhang et al. 1997b; 
(6) Gierli\'nski et al. 1997a; 
(7) Cui et al. 1996; 
(8) Belloni et al. 1996; 
(9) Dotani et al. 1996; 
(10) Phlips et al. 1996; 
(11) Zdziarski et al. 1997; 
(12) Gierli\'nski et al. 1997b }
\end{table}

    Certain of these figures deserve special comment.  $\Ls$ in the hard
state is particularly uncertain.  In works of Ba{\l}uci\'nska-Church et al.
(1995) and Ebisawa et al. (1996), the spectrum in
the {\it ROSAT} band is fit by a sum of a blackbody and a power-law.  They take
$\Ls$ to be only the part due to the blackbody; we believe that
it is better described by the total of the two, and therefore find a
soft luminosity two times greater than they do. 
In the soft state, $C$ is very difficult to determine, 
since the amount of reflection  
depends on the assumed run of ionisation with radius 
(Chris Done, private communications). 
We will point out the degree to which
our conclusions are sensitive to this uncertainty.

   Some of the physical parameters of the system may be derived (or
at least constrained) almost directly from observables.  For example,
the electron temperature in the corona (measured in electron rest mass
units) is very closely related to the cut-off
energy of the hard component: $\Theta \simeq f_x E_c/(m_e c^2)$, where
$f_x$ is a number slightly less than unity, whose exact
value depends on $\taut$.  Similarly,
$L_h \approx \Lh\left( 1+C4\pi da/d\Omega\right) /(1 + C)$, 
where $da/d\Omega$ is the albedo per unit solid angle for Compton 
reflection, and we assumed that the hard component has an isotropic 
angular distribution and $\taut\ll 1$. 

The intrinsic disc luminosity is
\begin{equation}
\Lsintr = \Lstot
- \frac{C L_h}{1+C} \left[1 - \int \, d\Omega \, da/d\Omega \right],
\end{equation}
where 
$\Lstot=\Ls/p(\theta)\left[1 - D\left(1 - e^{-\taut}\right)\right]$
is the total soft luminosity, and 
$p(\theta)$ gives the angular distribution of the disc radiation 
(normalized so that $\int \, d\Omega \, p(\theta) = 4\pi$). 
Taking the disc emission to be approximately black body, its inner radius is
\begin{equation}
 R_{s} = \left( {\Lstot \over 4\pi \sigma T_{s}^4 }\right)^{1/2}
 \left( {\fcol^2 \fGR^2 \over f_m}\right), 
\end{equation}
where $T_s$ is the effective temperature at the inner edge.   The additional
correction factors are: $\fcol$, the ratio between
the local color temperature and the local effective temperature;
$\fGR$, which incorporates the general relativistic corrections linking 
the local color temperature to the observed one; and $f_m$, which accounts
for a proper integration over the disc's surface brightness distribution.  
For estimates of the correction factors see, e.g., Shimura \& Takahara
 (1995) and Zhang et al. (1997a). 
The combination of all these correction factors is 
likely to be close to unity, but is
uncertain at the 50\% to factor of 2 level.

    We next employ the two following analytic scaling approximations for
thermal Comptonization spectra found by Pietrini \& Krolik (1995): 
$D(1 + S) = 0.15 \alpha^4 L_h / L_{s}^{\rm tot}$
and
$\taut = 0.16 / ( \alpha \Theta)$.
The first expresses how the power-law hardens as the heating rate in the
corona increases relative to the seed photon luminosity; the second
expresses the trade-off in cooling power between increasing optical depth
and increasing temperature.  

\begin{figure}
\begin{center}
\leavevmode
\epsfxsize=8.4cm \epsfbox{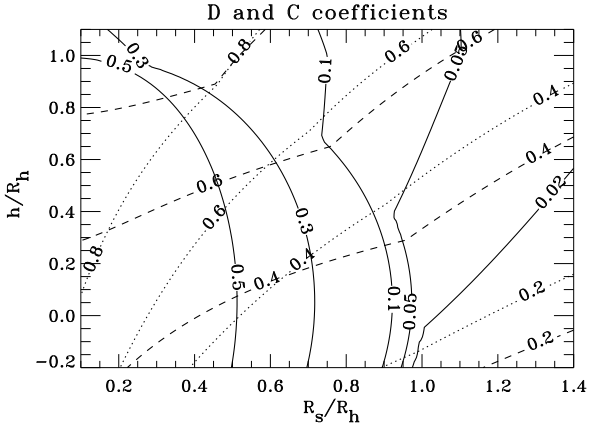}
\end{center}
\label{fig:dc}
\caption{Coefficients $D$ and $C$ as function of $h/R_h$ and 
$R_s/R_h$. Solid  curves represent 
contours of the constant $D$.  Dashed curves give contours of constant 
$C$ for $\taut=1.3$ (corresponding to the HS). Dotted curves 
give $C$ for $\taut=0.3$ corresponding to the SS. 
} 
\end{figure}

Finally, we assume a description of the geometry involving the minimum
number of free parameters consistent with describing a reasonable universe
of possibilities.   We take the disc to be an annulus of inner radius $R_s$
extending to infinite radius, with infinitesimal vertical thickness.
We similarly assume that the corona is a pair of spheres uniform inside, each
of radius $R_h$, and centered a distance $h$ from the disc midplane
(when $h = 0$, the two collapse into a single sphere).
This simplified description should be an adequate qualitative stand-in
for any coronal geometry which is unitary (i.e. not broken into many
pieces), axisymmetric, and reflection symmetric about the disc plane.
Both $C$ and $D$ may then be written in terms of $R_s/R_h$ 
and $h/R_h$.
We computed $C$ taking into account 
electron scattering of the reflected radiation in the corona and 
assuming that the reflection emissivity is $\propto (r^2+\hbar^2)^{-3/2}$
(where $\hbar$ is the averaged height of the corona above the disc plane) 
and its angular distribution is $\propto \cos\theta$. 
In computations of $D$, we assumed that the disc surface brightness 
is $\propto r^{-3}$ and its intensity distribution is isotropic, i.e.,
$p(\theta) = 2 \cos\theta$.
 Contour plots of the coefficients $C$ and $D$ are presented in 
Figure~\ref{fig:dc}. From  $C$ and $D$ we can determine both $R_s/R_h$ and 
$h/R_h$.

   With these relationships and assumptions, we arrive at the derived
quantities set out in Table 2.  Although they are based on approximate 
expressions (and setting $4\pi da/d\Omega = 0.3$, along with
$p = \fcol^2 \fGR^2/f_m = 1$), we will
show in the following section that they are confirmed with only small
quantitative corrections by detailed calculations.

\begin{table}
\label{tab:2}
\caption{Inferred properties of Cyg X-1 
in its soft and hard states}
\begin{tabular}{llll}  
\hline
{Parameter} & {Hard state} & { } & {Soft state} \\
$\Theta$$^{\rm a}$ &   $\sim$0.2 &     &     $\simgreat$ 0.3   \\
$\Lsintr/L_h$$^{\rm b}$ & $\simless 0.1$ & & $\sim 3$  \\
$R_s$$^{\rm c}$                            & 500 km & & 100 km \\
$D(1+S)$$^{\rm d}$ & 0.08 &  &  0.3     \\
$\taut$$^{\rm e}$   &  $\simeq 1.3$       & &  $\simless 0.35$ \\
$R_s/R_h$$^{\rm f}$ &  $\simeq$ 0.9    & &  $\simeq 0.65$  \\
$h/R_h$$^{\rm g}$ &  $\simeq 0.2$  & &  $\simeq 0.5 $   \\
$l_h$$^{\rm h}$    & $\simeq 20$  & & $\simeq 15$ \\  
\hline
\end{tabular} 

$^{\rm a}${Electron temperature, $\Theta\equiv k\Te/mc^2$} \\
$^{\rm b}${Ratio of intrinsic soft luminosity to the intrinsic hard
luminosity (all the 
correction factors are here assumed to
combine to unity)} \\
$^{\rm c}${Inner radius of the cold disc} \\
$^{\rm d}${Covering fraction of  ``corona" around thermal disc times
the seed photon correction factor} \\
$^{\rm e}${Radial Thomson optical depth of the hot cloud} \\
$^{\rm f}${Inner radius of the disc relative to the hot cloud} \\  
$^{\rm g}${Elevation of the corona above to the disc relative to the size 
of the hot cloud} \\ 
$^{h}${Hard compactness, $l_h \equiv L_h/R_h (\sigmat/m_ec^3)$}
\end{table}

We can determine from Figure~\ref{fig:dc} that in the hard state 
$h/R_h \simeq 0.2$. This corresponds to an almost spherical corona.
The cold disc penetrates only  a short way into the corona 
(see Gierli\'nski et al. 1997a and Dove et al. 1997  for a similar 
conclusion). If the disc were filled, relatively small $C$ could be 
achieved for large $\tau$ due to electron scattering of the reflection 
component, but it would be difficult to get $D$ small. The same argument
also holds if the corona is larger than the disc (it would be easy 
to find geometries in which $C$ is small, but then it becomes more 
difficult for $D$ to also be small). 
In the soft state, $h/R_h\approx 0.5$, so that the corona is
slightly elongated along  the axis of the system (an outflow?).
Now, the corona covers a significant 
fraction of the inner disc ($R_s/R_h\approx0.6$).

There is, of course, some uncertainty in determining  
$C$ and $D$ from observations. $C$ is usually  known with $\sim 25\%$ accuracy. 
This translates into similar uncertainty in $R_s/R_h$, and 
$\sim 50\%$ uncertainty in $h/R_h$. 
Estimations of $D$ that are uncertain within a factor of 2 (mostly 
due to the strong dependence on the spectral index $\alpha$) 
translate to $\sim 30\%$ and $\sim 15\%$ uncertainty in $R_s/R_h$ 
for the SS and HS, respectively, while $h/R_h$ is almost unaffected.

\section{DETAILED CALCULATIONS}

    To verify these approximate arguments and refine our estimates of
the inferred system parameters, we have computed exact numerical
equilibria for a variety of parameters. 
We solved energy balance and electron-positron 
pair balance equations coupled with the radiative transfer. 
This was done using the code described in Poutanen \& Svensson (1996). 
We assumed that the corona is spherical and centered 
on the black hole (i.e. $h=0$), based on the fact that 
in both states the derived value of $h/R_h$ is significantly smaller than 1. 

    We divided these calculations into two sets, one designed to mimic
the hard state, the other the soft.  All but one of the system parameters
needed to define an equilibrium are given by the analytic estimates of
the previous section.  The one exception is $\taup$, the net lepton
optical depth.  Although the total optical depth $\taut$
is reasonably well determined by observations, $\taup$ is not, because
there is no easy way to tell from observations what portion of the
Compton opacity is due to $e^{\pm}$ pairs.  In terms of the ratio $z \equiv
n_+/n_p$, $\taut = \taup (2z + 1)$.  The one free parameter of
the calculations is therefore $\taup$. 

\begin{figure}
\begin{center}
\leavevmode
\epsfxsize=8.4cm \epsfbox{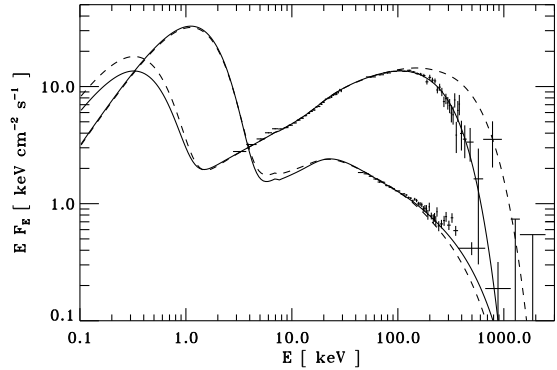}
\end{center}
\label{fig:fit}
\caption{Theoretical spectra predicted by the calculations described in \S~3.
In both spectral states, the absolute normalization of the luminosity is
arbitrary.  The solid curve for the hard state corresponds to model HS3,
and the dashed curve to model HS2 (inclination of the disc 
is assumed to be 30$^{\rm o}$).  
The solid curve for the soft state
corresponds to model SS2, and the dashed curve to model SS3.  The
discrepancies between predictions and observations, particularly at the high
energy end, for HS2 and SS3 show that there cannot be too many pairs in
the hard state, or too few in the low state.  The hard state data are from 
Gierli\'nski et al. (1997a) and the soft state data are publically 
available from the HEASARC. }
\end{figure}

In the hard state models, 
we fixed $l_h = 20$, $T_s = 0.13$~keV, $\Lsintr/L_h =
0.1$, and $R_s/R_h = 0.9$, and let $\taup = 0$, 1.0, and 2.0 (models 
HS1, HS2, HS3, respectively). Calculations gave 
({$\taut$}, {$\Theta$}, {$z$})= (0.87, 0.39, $\infty$) for model HS1, 
(1.12, 0.31, 0.06) for HS2, and (2.001, 0.16, 0.0002) for HS3. 
Selected predicted spectra are
compared with observations in Figure~\ref{fig:fit}. 
When $\taup \ll 1$, pairs dominate,
and $\Theta \simeq 0.4$; $\taup \simgreat 1$ leads to $z \ll 1$, and
the greater optical depth depresses the electron temperature.  
  The best agreement with the observed spectrum (and the
quality of this agreement is actually quite good) is found
with $\taut \simeq \taup = 2$, {\it i.e.}, a state in which the pair
contribution is negligible (assuming that pairs are thermal). 
Thus, the approximate parameters of Table 2
come very close both to a truly self-consistent equilibrium, and to
producing an output spectrum like the one seen.

 In the soft state models we fixed $l_h = 15$, $T_s = 0.4$~keV,
$\Lsintr/L_h = 3$, and $R_s/R_h = 0.65$, and let $\taup =$ 0, 0.1,
and 0.3 (models SS1, SS2, SS3, respectively).  
We find the following results: ({$\taut$}, {$\Theta$}, {$z$})  = 
(0.24, 0.37, $\infty$) for model SS1, (0.25, 0.37, 0.75) for SS2, and 
(0.32, 0.31, 0.03) for SS3.

In many respects, the soft state is quite similar to the hard state; it
differs primarily in having roughly an order of magnitude smaller optical
depth, and in the much greater importance of intrinsically generated soft
photons.  These changes do not alter very much the critical $\taup$
below which pairs dominate: it is still a few tenths.  However, in
contrast to the hard state, the best agreement with observations (and this
fit is again quite satisfactory) is found with $z \simgreat 1$.  Again
we see that the analytic arguments are reasonably good guides to the results
of more exact calculation.

\section{DISCUSSION}

   These inferences allow several strong conclusions to be made.  Most
importantly, the fact that such a self-consistent picture may be drawn gives
additional support to the basic picture of thermal Comptonization for the origin
of the hard X-rays.  This suggestion was part of the original proposal of
Shapiro et al. (1976), who suggested that in the hard state, 
the inner portion of
the accretion disc puffs up into a hot Comptonizing corona.  Their picture
of the geometry is also confirmed.  In addition, we now see that in the
soft state the disc extends inward almost as far as it can, while the
corona surrounds it out to several gravitational radii.

    We are also now able to see that the disc receives only a
minority of the dissipation in the hard state, but is the site of most of
the heat release in the soft state.  This conclusion is especially robust,
for it depends primarily on the single ratio $\Lh/\Ls$.  If we have
overestimated $\Ls$ in the hard state, the fraction of the dissipation
taking place in the disc becomes even smaller.

The radius of the disc's inner edge
shrinks by roughly a factor of 5 between the hard state and the soft state.
This ratio is somewhat sensitive to several uncertainties.  First, $R_s \propto
\left[\Ls\right]^{1/2}$, and second,
$R_s$ in the soft state can be affected by uncertainty in the scaling 
factors $p(\theta)$, $\fcol$, $\fGR$, and $f_{m}$.  In terms of gravitational 
radii $r_g = GM/c^2$, $R_s \simeq 40 r_g M_{10}^{-1}$
in the hard state, and $\simeq 8 r_g M_{10}^{-1}$ in the soft state.  Here
$M_{10}$ is the mass of the black hole scaled to $10 M_{\odot}$, the best
estimate of Herrero et al. (1995).

    Estimates of $R_s$, combined with the intrinsic luminosity of the disc, 
also permit us to place strong bounds on the mass accretion rate through
the disc, $\dot M_d$.  Almost independent of the spin of the black hole,
mass passing through a thin disc extending outward from the $\simeq 40 r_g$
inferred for the hard state has a net binding energy $\simeq 0.02$ of its
rest mass.  The intrinsic disc luminosity, at least, must be drawn from this
store of energy; $\dot M_d$ in the hard state must then be at least $0.4 \times
10^{-8} M_{\odot}$~yr$^{-1}$ (this estimate is sensitive to uncertainty in
$\Ls$; it scales roughly $\propto \left[\Ls\right]^{3/2}$).  The greatest
possible $\dot M_d$ would be required if all the energy for coronal heating is
taken from matter accreting through the disc (that is, inside $R_s$, the
specific angular momentum of the matter changes, but not its specific energy). 
 That upper bound is $\dot M_d \simeq 4 \times 10^{-8} M_{\odot}$~yr$^{-1}$.
In the soft state, the efficiency of the disc varies from $\simeq 0.05$ (zero
spin) to $\simeq 0.1$ (maximum rotation).  Similar reasoning then gives a lower
bound of 0.7 -- $1.4 \times 10^{-8} M_{\odot}$~yr$^{-1}$ and an upper
bound only about 20\% greater (because nearly all the luminosity in the
soft state is radiated by the disc).  If $\dot M_d$ in the hard state is
near the lower bound, the efficiency of the inner
corona in the hard state must be $\simeq 0.2$ in order to explain the
total luminosity.  If so, the black hole would have to be rotating
relatively rapidly.
 
   It is also of interest to compare $\dot M_d$ to the accretion rate which
would produce an Eddington luminosity if radiation were created at
the efficiency of a maximal Kerr black hole---$6 \times 10^{-8} M_{10}
M_{\odot}$ yr$^{-1}$.  In those units, the estimates of the previous
paragraph give normalized accretion rates $\dot m = 0.07$ (or possibly
as high as $\dot m = 0.7$) in the
hard state, and $\simeq 0.1$ -- 0.2 in the soft state.

Meanwhile, the corona shrinks in radial scale by only half as much as
the disc when Cyg X-1 moves from the hard state to the soft.  In the
latter state, it covers a sizable portion of the inner disc.
Despite the sharp change in size and luminosity, the compactness of
the corona changes by no more than a factor of a few (incorporating the
uncertainty in $R_s$) as a result of the hard-soft transition.  However,
its optical depth drops by an order of magnitude.  
Although
there are few $e^{\pm}$ pairs in the equilibrium characterizing the hard
state, they can be comparable in number to the net electrons in the soft
state.  This fact also means that the fall in $\taup$ from the hard
state to the soft is even greater than the fall in $\taut$.

 We close by noting that in the hard state there must be a hole in
the central region of the thermal disc which is much bigger 
than the radius of the innermost stable orbit.  
The origin of this hole is a question of prime importance.  One possibility
arises from the fact that the disc becomes dominated by radiation pressure
inside $\simeq 35 (\dot m/0.1)^{2/3}$ gravitational radii.
As shown by Shakura \& Sunyaev (1976),
in the strict $\alpha$-model, radiation pressure-dominated discs are thermally
unstable.  It is interesting that the inner radius of the
disc in the hard state corresponds quite closely to this possible point
of instability.  We are then left with the question of how
a thin disc is able to persist in to much smaller radius in the soft
state, when $\dot m$ is still $\sim 0.1$.  
Another possibility is that the inner disc in the hard
state is advection-dominated, while in the soft state the accretion rate is
just a bit too high for that to happen \cite{ny95}.   While the
creation of an advection-dominated region may be a nonlinear end-state
of the thermal instability, so that this model provides an attractive
explanation for the hard state, it also leaves open how the thin disc
survives thermal instability in the soft state.

\section*{ACKNOWLEDGMENTS}

     J.H.K. would like to thank Roland Svensson and the Stockholm
Observatory for their hospitality during his visit when this project was begun.
He was partially supported by NASA Grant NAGW-3156.
J.P. and F.R. would like to acknowledge the hospitality that they, in their 
turn, received at the Department of Physics and Astronomy of Johns Hopkins 
University. 
J.P. was supported by a grant from the
Swedish Natural Science Research Council and F.R. by a grant from the
Royal Swedish Academy of Sciences. 
The authors  would also like to thank Chris Done, the referee, 
for helpful suggestions.

\end{document}